\documentclass[aps,prc,twocolumn,floatfix,superscriptaddress]{revtex4}
\usepackage{epsfig}
\usepackage{amsmath}
\usepackage{color}

\usepackage{graphicx}

\definecolor{blue}{rgb}{0.05, 0.05, 0.5}

\begin{document}

\title{Medium Modification of Charm Production in Relativistic Heavy Ion Collisions due to Pre-equilibrium Dynamics of Partons 
at $\sqrt{s_{\textrm {NN}}}$= 0.2--5.02 TeV}
\author{Dinesh K. Srivastava}
\email{dinesh@vecc.gov.in}
\affiliation{Variable Energy Cyclotron Centre, HBNI, 1/AF, Bidhan Nagar, Kolkata 700064, India}
\author{Rupa Chatterjee}
\email{rupa@vecc.gov.in}
\affiliation{Variable Energy Cyclotron Centre, HBNI, 1/AF, Bidhan Nagar, Kolkata 700064, India}

\begin{abstract}
We study the production dynamics of charm quarks in the parton cascade model for relativistic
heavy ion collisions at RHIC and LHC energies. The model  is 
eminently suited for a study of the pre-equilibrium dynamics of charm quarks at modest transverse momenta. The treatment
is motivated by QCD parton picture and describes the dynamics of an ultra-relativistic heavy-ion collision in terms of
 cascading partons which undergo scattering and multiplication while propagating. 
We find considerable suppression of charm quarks production in $AA$ collisions compared to those for $pp$
collisions at the same $\sqrt{s_{\text {NN}}}$ scaled by number of collisions. 
This may be important for an accurate determination of
  energy loss suffered by charm quarks while 
traversing the thermalized quark gluon plasma. 
\end{abstract}


\maketitle

\maketitle
\section{Introduction} 

The existence of Quark Gluon Plasma (QGP), a deconfined strongly interacting matter, which
 was predicted by lattice Quantum Chromo Dynamics  calculations (see, e.g, 
Ref~\cite{Ratti:2018ksb} for a recent review), is now routinely produced in relativistic heavy ion collisions at BNL Relativistic Heavy Ion Collider and CERN Large Hadron Collider.  It is believed to have filled the early universe till a few micro-seconds after the Big Bang. The study of QGP
has remained one of the most rewarding disciplines of modern nuclear physics for more than three decades.

 The observation of large elliptic
 flow of hadrons~\cite{Adler:2002ct, Adler:2003kt} and jet-quenching~\cite{Adcox:2001jp, Adams:2003kv}, because of the energy loss of high energy partons in the hot and dense medium, are the most prominent early signatures of the formation of QGP in these collisions. Additional confirmation has been provided by detection of thermal radiation of photons form these collisions~\cite{Aggarwal:2000th, Adare:2008ab, Adam:2015lda}.
The unexpected surprises have been provided by the parton recombination as a mechanism of hadronization~\cite{Fries:2003vb}
and the very low
viscosity (see e.g., Ref~\cite{Song:2007ux, Luzum:2008cw, Song:2010mg}). 

Coming years will explore its properties to a great deal of accuracy and once the Facility for Anti-proton and Ion Research, Darmstadt (FAIR) and the Nuclotron-based Ion Collider Faсility, Dubna (NICA) start working, the next frontier of this fascinating field, namely QGP at high baryon density and low temperatures, which perhaps forms the core of large neutron stars~\cite{Annala:2019puf}, will be open for exploration. The Future Circular Collider (FCC) will provide an opportunity to study $pp$ and $AA$ collisions at unprecedented high centre of
mass energies~\cite{Dainese:2016gch,Armesto:2016qyo,Armesto:2014iaa}.

Charm quarks have emerged as a valuable probe of the evolution dynamics of quark gluon plasma. This was realized quite early in the literature. The large mass of charm quarks ensures that they are produced
only in processes involving a sufficiently large $Q^2$. This makes these interactions amenable to perturbative QCD calculations. The later interactions conserve charm and the hadrons containing charm are easily identified. More than three decades ago, 
Svetistky~\cite{Svetitsky:1987gq} obtained the drag and diffusion coefficients for them by considering that they execute a Brownian motion in the QGP. A first
estimate of the radiative energy loss of heavy quarks was also obtained by authors of Ref.~\cite{Mustafa:1997pm}
using some simplifying assumptions. These early studies have been brought to a very high degree of sophistication by now.
The energy loss suffered by heavy quarks due to scatterings and radiation of gluons have been been estimated and its consequences 
have been explored in detail (see, e.g.
\cite{GolamMustafa:1997id, vanHees:2004gq, Moore:2004tg, vanHees:2005wb, Peigne:2008nd, Gossiaux:2008jv, Gossiaux:2009mk, He:2011qa, Cao:2011et, Abir:2012pu, Meistrenko:2012ju, Alberico:2013bza, Berrehrah:2014kba,Song:2015sfa, Song:2015ykw, Horowitz:2015dta, Nahrgang:2016lst, Scardina:2017ipo, Plumari:2017ntm, Sheikh:2017ock}). The temperature dependence of the drag coefficient has also been calculated using lattice QCD~\cite{Banerjee:2011ra, Ding:2012sp, Kaczmarek:2014jga}. A phenomenological extension of Wang, Huang, Sarcevic model~\cite{Wang:1996yh, Wang:1996pe} was used by authors of Ref.~\cite{Younus:2012yi} to extract energy loss of
charm quarks in the quark gluon plasma and their azimuthal correlations~\cite{Younus:2013be} and a comparative study of different energy loss mechanisms for heavy quarks
at central and forward rapidities was performed by authors of Ref.~\cite{Jamil:2010te}.

However, all the above studies mostly start with the assumption of a thermally and chemically equilibrated plasma at some
formation time $\tau_0 \approx$ 0.2--1.0 fm/$c$. The assumption of chemical equilibration has been relaxed in some studies for determination of the drag and diffusion~\cite{GolamMustafa:1997id}. The consequences of this relaxation for 
 interactions~\cite{Lin:1994xma,Younus:2010sx} following their initial production in prompt 
collisions~\cite{Levai:1994dx} have also been studied. The drag and diffusion co-efficients for heavy quarks for the pre-equilibrium phase have been studied by replacing the distributions of quarks and gluons by Colour Gluon Condensate model inspired distributions~\cite{Das:2015aga}. 

The thermalization of heavy quarks by assuming that they perform Brownian motion in a thermal bath of gluons was studied quite some time 
ago~\cite{Alam:1994sc}. Heavy quark thermalization and flow has been studied in considerable detail within a partonic transport model BAMPS (Boltzmann Approach of Multi Parton Scatterings) by the authors of 
Ref.~\cite{Uphoff:2010sh, Uphoff:2011ad}, where the initial distribution of charm quarks was sampled from {\tt {PYTHIA}}.

The parton cascade model proposed by Geiger and Muller~\cite{Geiger:1991nj} has been refined by
Bass, Muller, and Srivastava~\cite{Bass:2002fh} with improved implementation of the dynamics of the collision with several interesting insights and results. It was further extended to implement the production of heavy quarks~\cite{Srivastava:2017bcm}. A box-mode implementation was shown to provide an accurate description of energy loss suffered by charm quarks in QGP at a given temperature due to collisions and radiations~\cite{Younus:2013rja}.

Recently it was employed to study the transport dynamics of parton interactions in $pp$ collisions at the LHC 
energies~\cite{Srivastava:2018dye}. These studies are of interest because of the QGP like features observed in high multiplicity events of these collisions.
The studies reported in Refs.~\cite{Srivastava:2017bcm,Srivastava:2018dye} were performed by neglecting the Landau Pomeranchuk Migdal (LPM) effect, which results in enhanced parton multiplication.

Authors of Ref.~\cite{Srivastava:2018nfu} have reported results for charm production in $pp$ collisions with the accounting of LPM effect. 
Their results indicate that
$pp$ collisions at the higher LHC energies may lead to formation of a dense medium. This, in turn triggers a suppression of radiations (and parton multiplication) due to the LPM effect.  
However, it was reported that, even after these suppressions, multiple scatterings occur among the partons and the transverse momentum distribution of charm quark is rather sensitive to such scatterings. These calculations also provided a reasonably good description of the charm distribution
measured at LHC energies. The bottom quarks, on the other hand, due to their very large mass are not likely to be
produced in multiple scatterings after the initial collisions and were not
affected by this suppression~\cite{Chatterjee:2018ulb}, at least for $pp$ collisions. 

These considerations presage a considerable influence of LPM effect in $AA$ collisions. We aim to study this pre-equilibrium dynamics
for charm production in $AA$ collisions, in the present work.

We briefly discuss the details of our formulation in the next section, give our results in Sec. III, and conclusions in Sec. IV.

\begin{figure}
\centerline{\includegraphics*[width=8.0 cm]{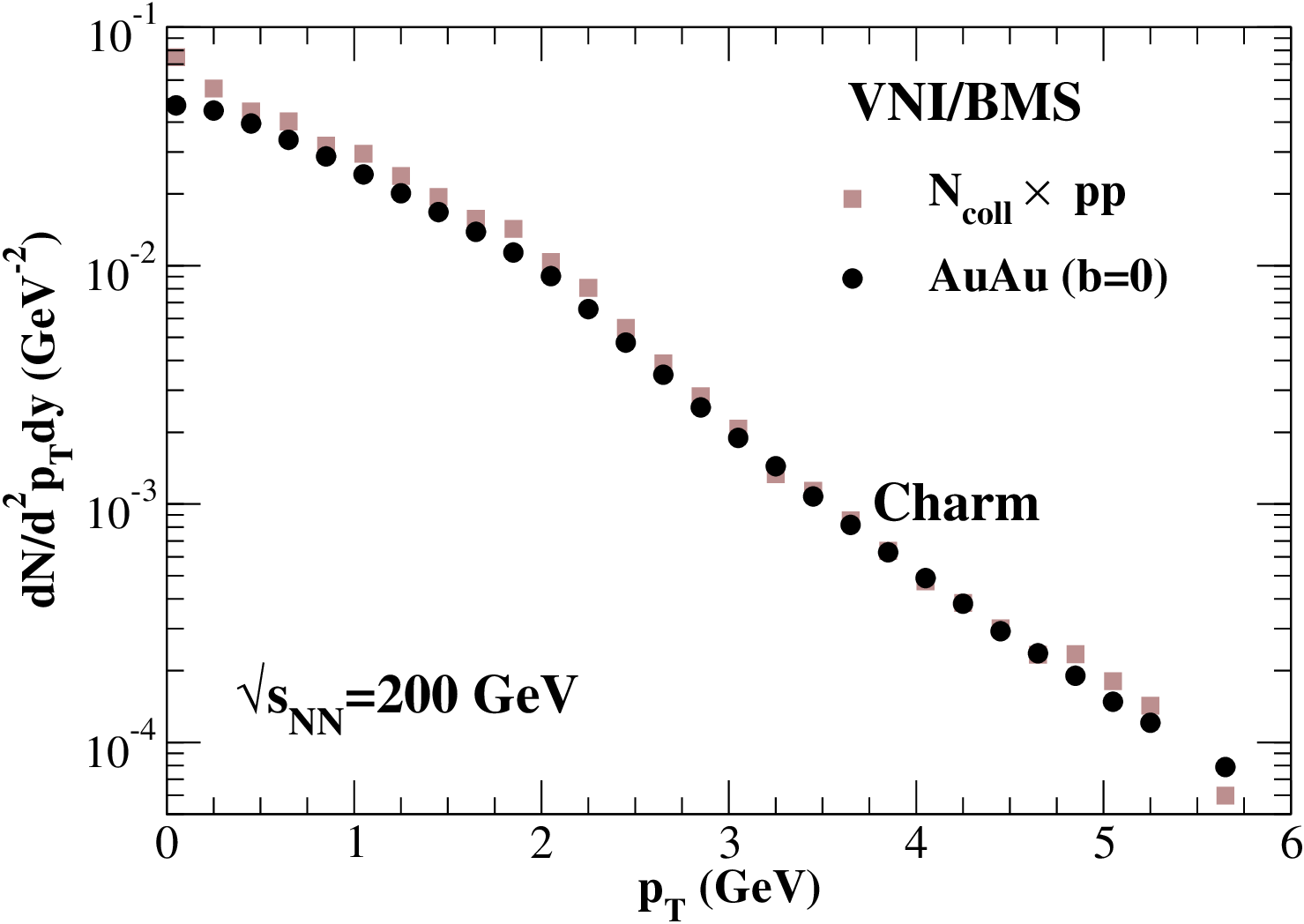}}
\centerline{\includegraphics*[width=8.0 cm]{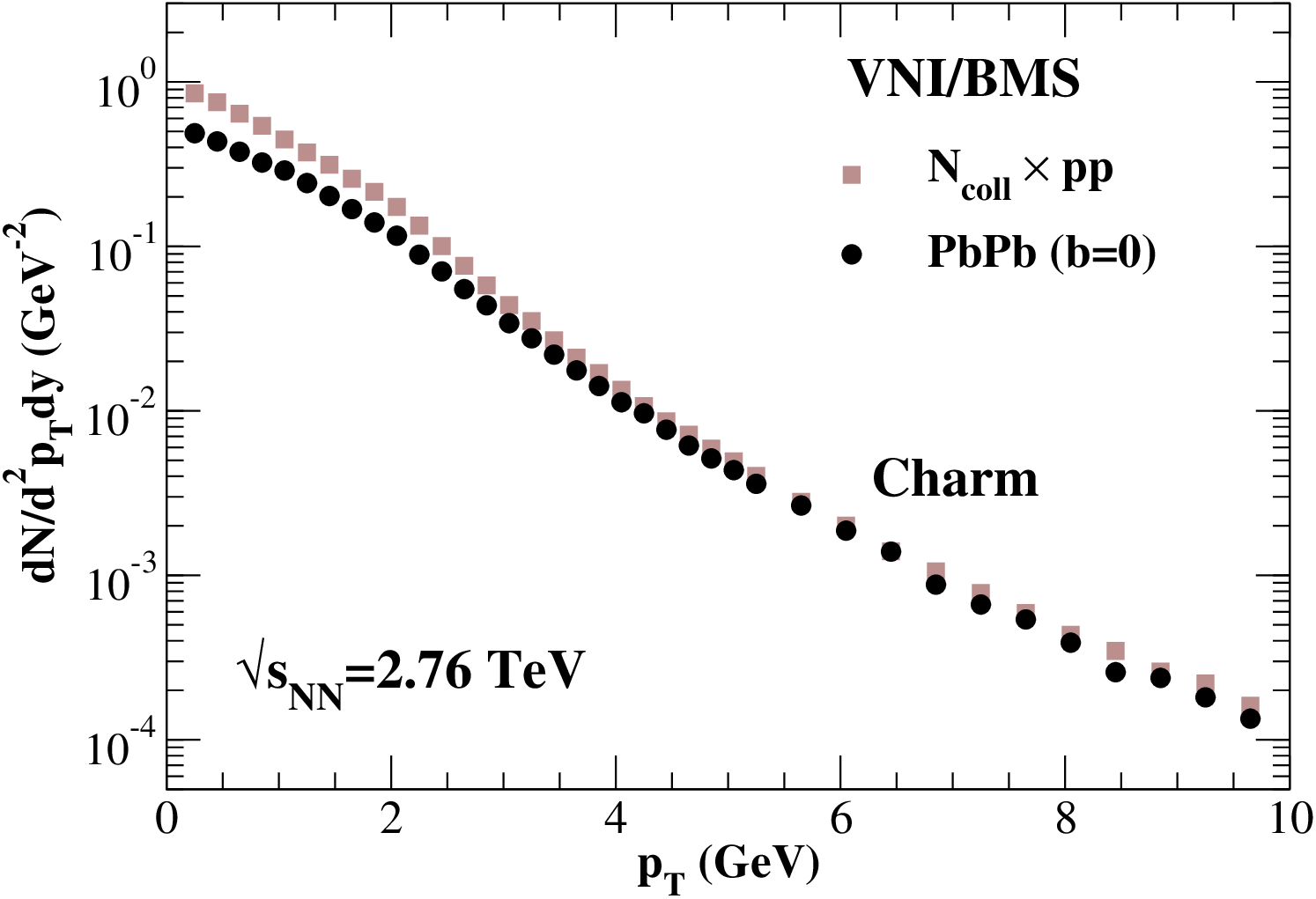}}
\centerline{\includegraphics*[width=8.0 cm]{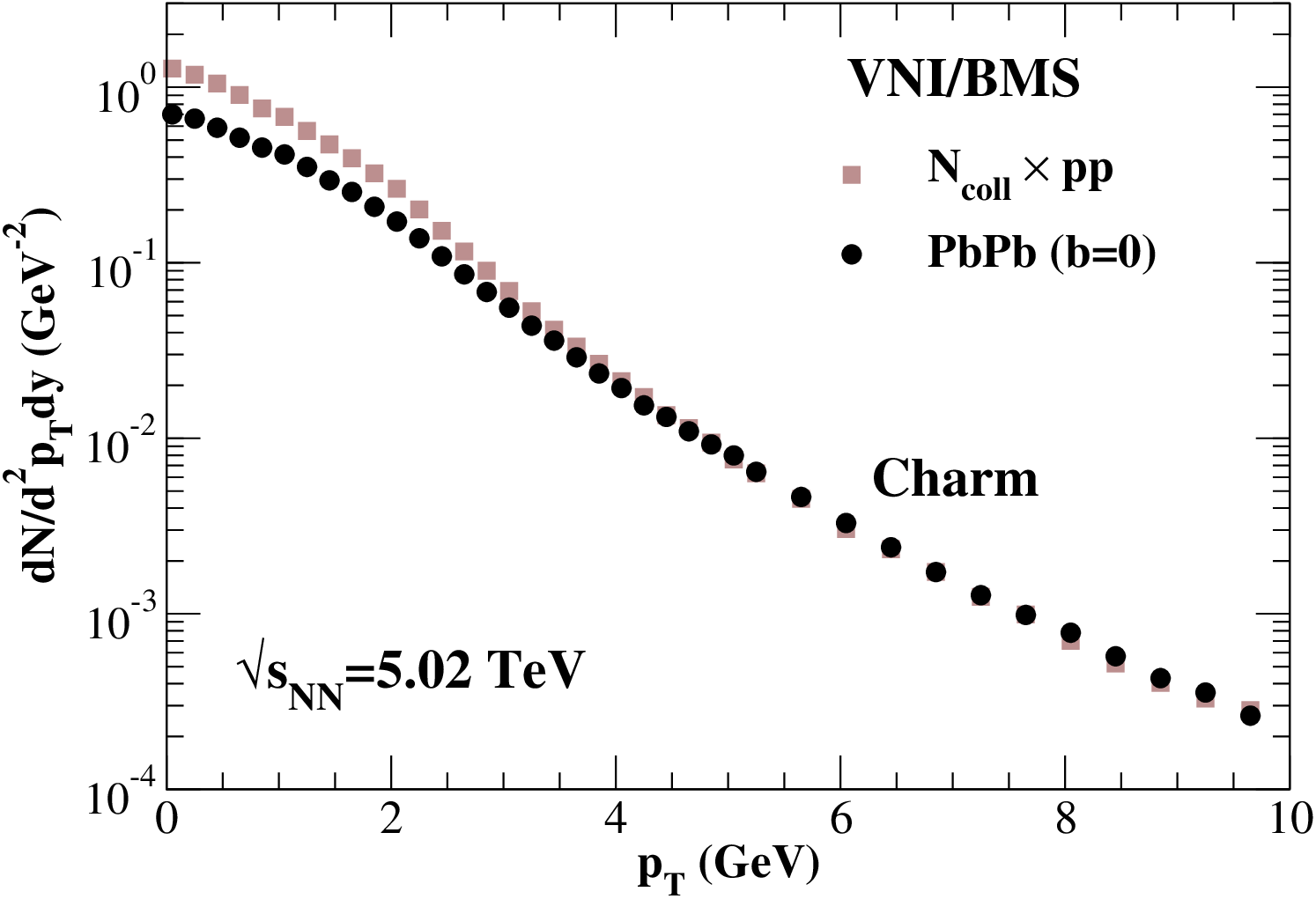}}
\caption{(Colour on-line) The $p_T$ distribution of charm quarks at the end of the pre-equilibrium phase for nucleus-nucleus collisions (for $b=0$ fm) 
and $pp$ collisions scaled by the number of collisions, at the same $\sqrt{s_{\text{NN}}}$ for $y=0$. 
 Results are given for Au+Au collisions at 200 AGeV (upper panel), Pb+Pb collisions at 2.76 ATeV (middle panel) and 5.02 ATeV (lower panel).}
\label{pt}
\end{figure}

\section{Formulation}

First of all, let us state the limited objective of our work in a little more detail. We realize that charm quarks, after their initial production in hard partonic scatterings will be a witness to the emergence of a thermally and possibly chemically equilibrated QGP followed by its expansion and cooling. These will see the beginning of the flow and hadronization. The heavy mesons containing charm quarks may also undergo scatterings during the hadronic phase. 
Thus these are valuable chroniclers of the system.

As mentioned earlier, the drag, diffusion, energy loss and flow experienced by them need to be understood in quantitative detail so that we can use them to determine the properties of the medium precisely. We realize that the charm quarks
will experience a considerable turmoil, before the thermally 
and chemically equilibrated plasma sets in at some formation time $\tau_0$ of the order of 0.2--1.0 fm/$c$. This suggests that we understand their dynamics {\it before} the system enters the so-called QGP phase, as some amount of medium modification of their momentum distribution could already happen by then. In absence of this the medium modification already sustained during the pre-equilibrium phase will have to be, per-force, 
accounted by adjusting the values for drag, diffusion and radiative energy loss during the QGP phase and later. 

The parton cascade model~\cite{Geiger:1991nj,Bass:2002fh} is eminently suited for this study for the following reasons. It starts from experimentally measured parton distribution functions and proceeds to pursue a Monte Carlo implementation of the Boltzmann equation  to study the time evolution of the parton density, due to semi-hard perturbative Quantum Chromo Dynamics (pQCD) interactions including scatterings and radiations within a leading log 
approximation~\cite{Altarelli:1977zs}. The $2\rightarrow 2$ scatterings among massless partons use the well-known matrix elements (see, 
e.g.~\cite{Owens:1986mp}) at leading order pQCD. The singularities present in the matrix elements are regularized by introducing a transverse momentum cut-off ($p_T^{\text{cut-off}}$ fixed at 2 GeV in the present work). 

The radiation processes 
($g\rightarrow gg$ and 
$q \rightarrow qg$) are, in turn, regularized by introducing a
virtuality cut-off, $\mu_0^i =\sqrt{m_i^2+ \mu_0^2}$, where $m_i$ is the current mass of quark (zero for gluons) and 
 $\mu_0$ is taken as 1 GeV. This is implemented using the well tested procedure implemented in 
${\tt {PYTHIA}}$~\cite{Sjostrand:2006za}. It has been reported earlier that the introduction of the LPM effect 
minimizes the dependence of the results on the precise value of $\mu_0$~\cite{Renk:2005yg,Srivastava:2018nfu}.  

The matrix elements for the 
$gg\rightarrow Q\overline{Q}$ and $q\overline{q} \rightarrow Q\overline{Q}$ processes do not have a singularity and the minimum $Q^2$ for them is $4M_Q^2$, which for charm quarks is more than 7 GeV$^2$ and amenable to calculations using pQCD. The $qQ\rightarrow qQ$ and $gQ\rightarrow gQ$ processes will require a $p_T^\text{cut-off}$ to avoid becoming singular, and it is taken as 2.0 GeV as before. The matrix elements for these are taken from Combidge~\cite{Combridge:1978kx}. For more details, the readers are referred to earlier publications~\cite{Srivastava:2017bcm}.

The scatterings among partons and radiation of gluons will lead to a rapid formation of a dense partonic medium for $AA$ collisions, even though the PCM involves only those partons which participate in the collisions leading to momentum transfers
larger than $p_T^\text{cut-off}$ and are radiated only their till virtuality of the mother parton
 drop to $\mu_0^i$ (see above). This would necessitate
introduction of LPM effect. We have already noted its importance for $pp$ collisions~\cite{Srivastava:2018nfu}.

We have implemented the LPM effect by assigning a formation time $\tau$ to the radiated particle:
\begin{equation}
\tau = \frac{\omega}{k_T^2},
\end{equation}
where $\omega$ is its energy and $k_T$ is its transverse momentum with respect
to the emitter. We further implement a scheme such that 
 during the formation time, the radiated particle does not interact. 
The emitter, though continues to interact and if that happens, the
radiated particle is removed from the list of participants and is thus excluded
from later evolution of the
system~\cite{Renk:2005yg} (see \cite{Srivastava:2018nfu}, for more detail). 

These aspects are incorporated in the Monte Carlo implementation of the
parton cascade model, {\tt {VNI/BMS}} which we use for the results given in the following. Before proceeding, we 
insist that PCM
does not include the soft scatterings which lead to flow etc. We have already mentioned that a good description of
charm production at LHC energies, using this procedure, was reported earlier~~\cite{Srivastava:2018nfu}.

\begin{figure}
\centerline{\includegraphics*[width=8.0 cm]{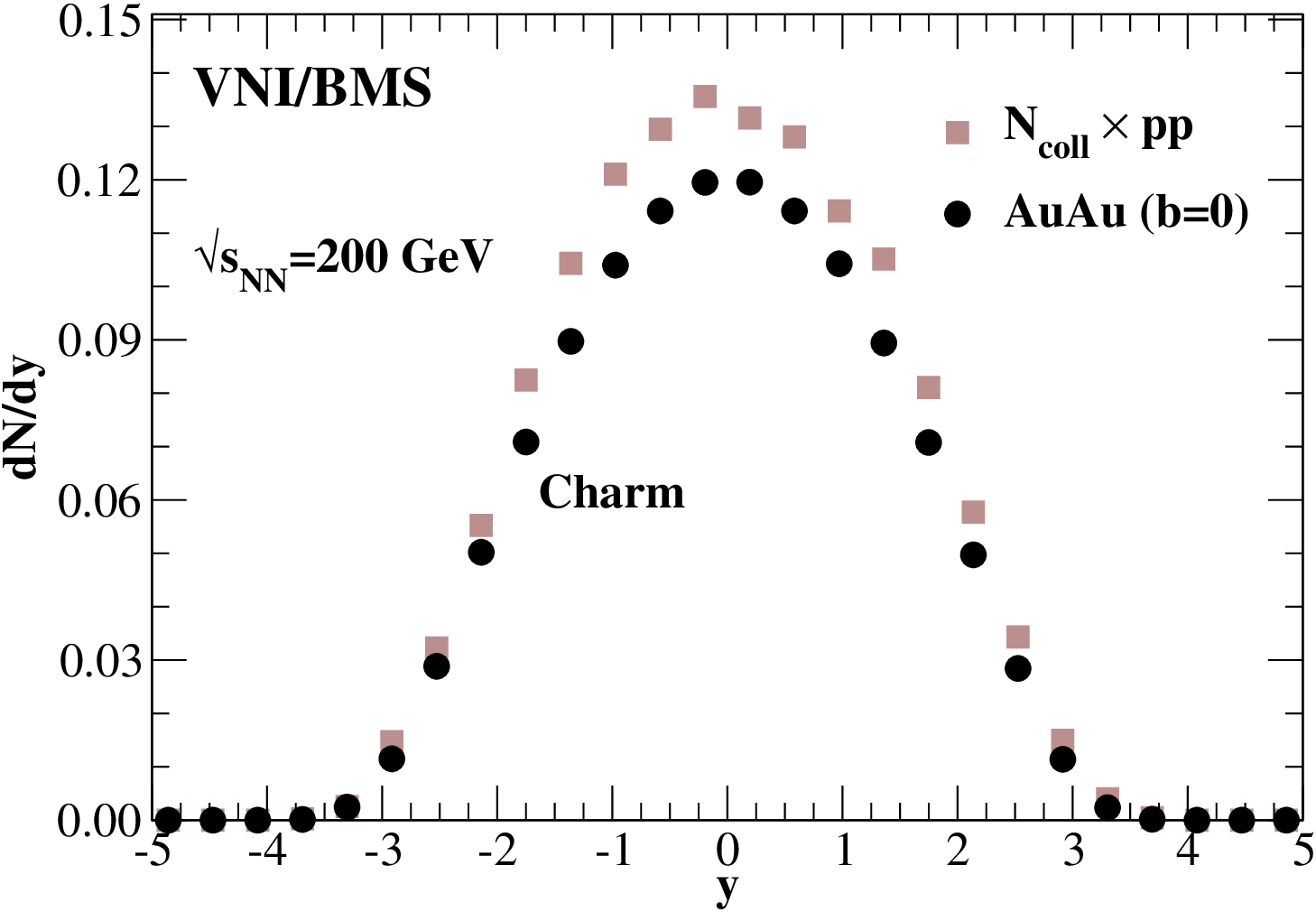}}
\centerline{\includegraphics*[width=8.0 cm]{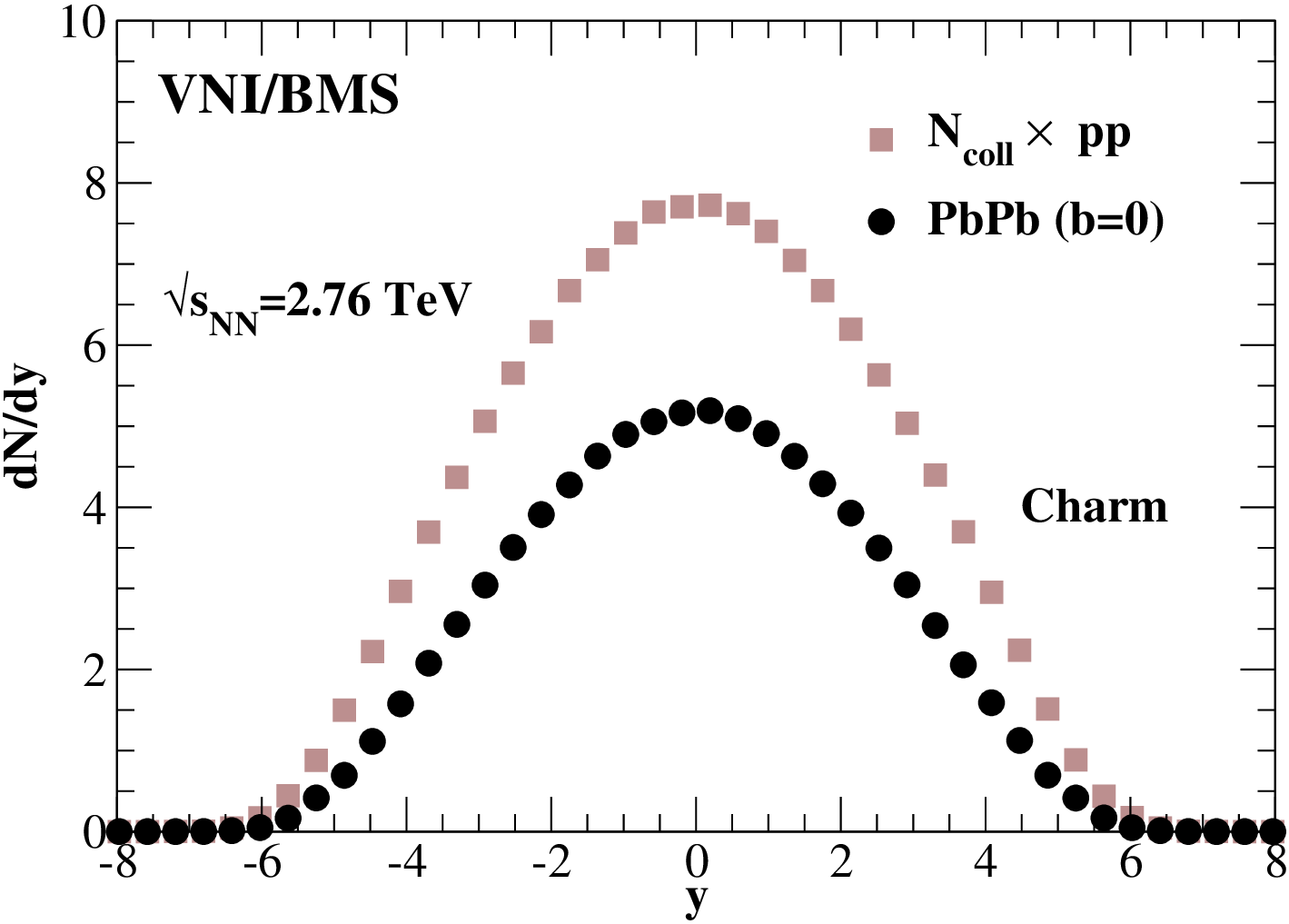}}
\centerline{\includegraphics*[width=8.0 cm]{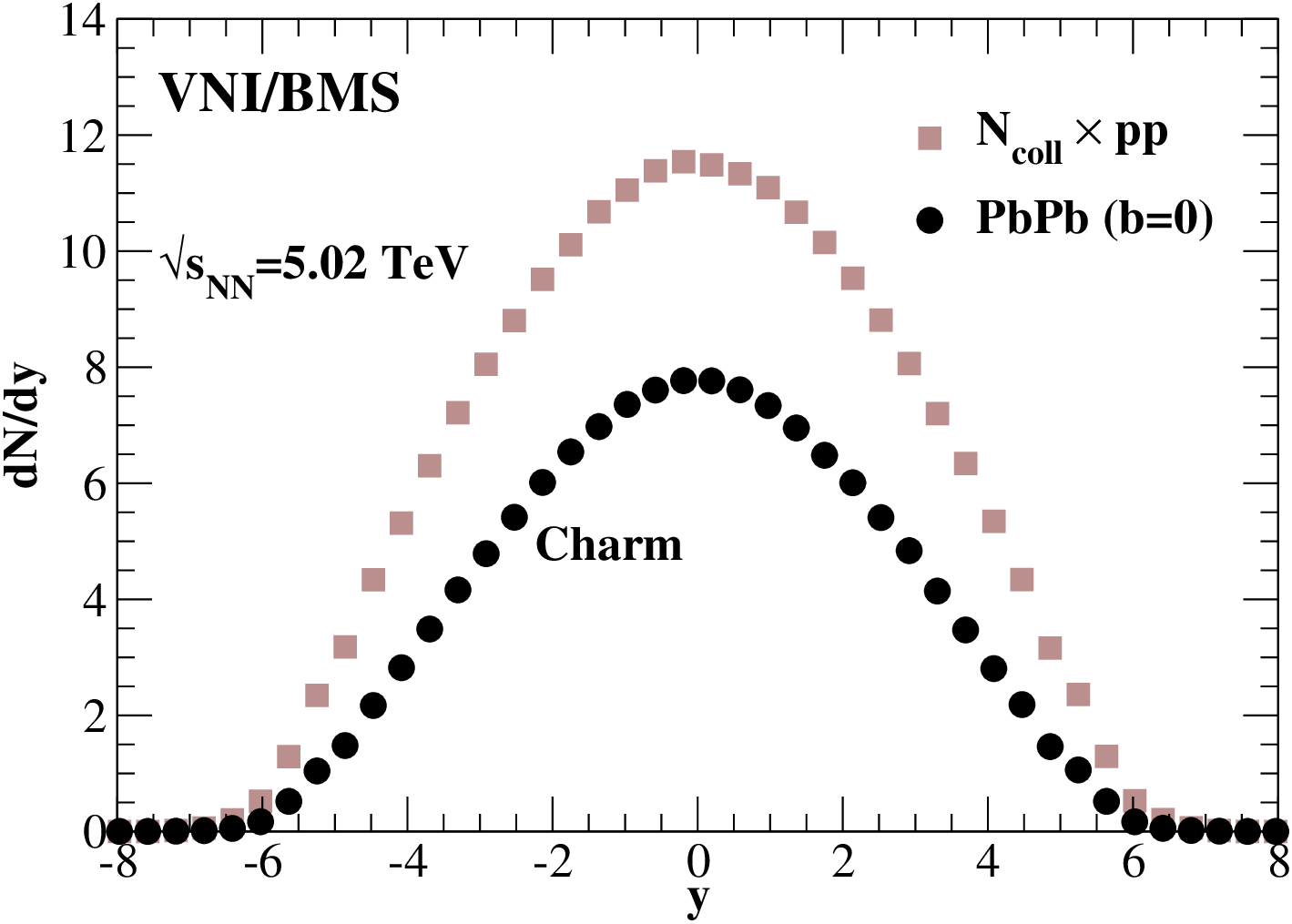}}
\caption{(Colour on-line) The $p_T$ integrated rapidity distribution of charm quarks at the end of the pre-equilibrium phase for nucleus-nucleus collisions (for $b=0$ fm) 
and $pp$ collisions scaled by the number of collisions, at the same $\sqrt{s_{\text{NN}}}$. 
 Results are given for Au+Au collisions at 200 AGeV (upper panel), Pb+Pb collisions at 2.76 ATeV (middle panel) and 5.02 ATeV (lower panel).}
\label{dy}
\end{figure}

\section{Results}

We have calculated production of charm quarks for $Au+Au$ collisions at 200 AGeV and for $Pb+Pb$ collisions at 
2.76 ATeV and 5.02 ATeV for zero impact parameter. Results for $pp$ collisions at the same centre of mass energy have also been included
for a comparison and for estimating the medium modification factor $R_{\text{AA}}$, such that
\begin{equation}
R_\textrm{AA}(p_T)=\frac{dN_\textrm{AA}/d^2p_T dy}{N_\text{coll} \times dN_\text{pp}/d^2p_T dy}
\end{equation}
where $N_\text{coll}$ is the number of binary nucleon-nucleon collisions for the given centrality.

\begin{figure}
\centerline{\includegraphics*[width=8.0 cm]{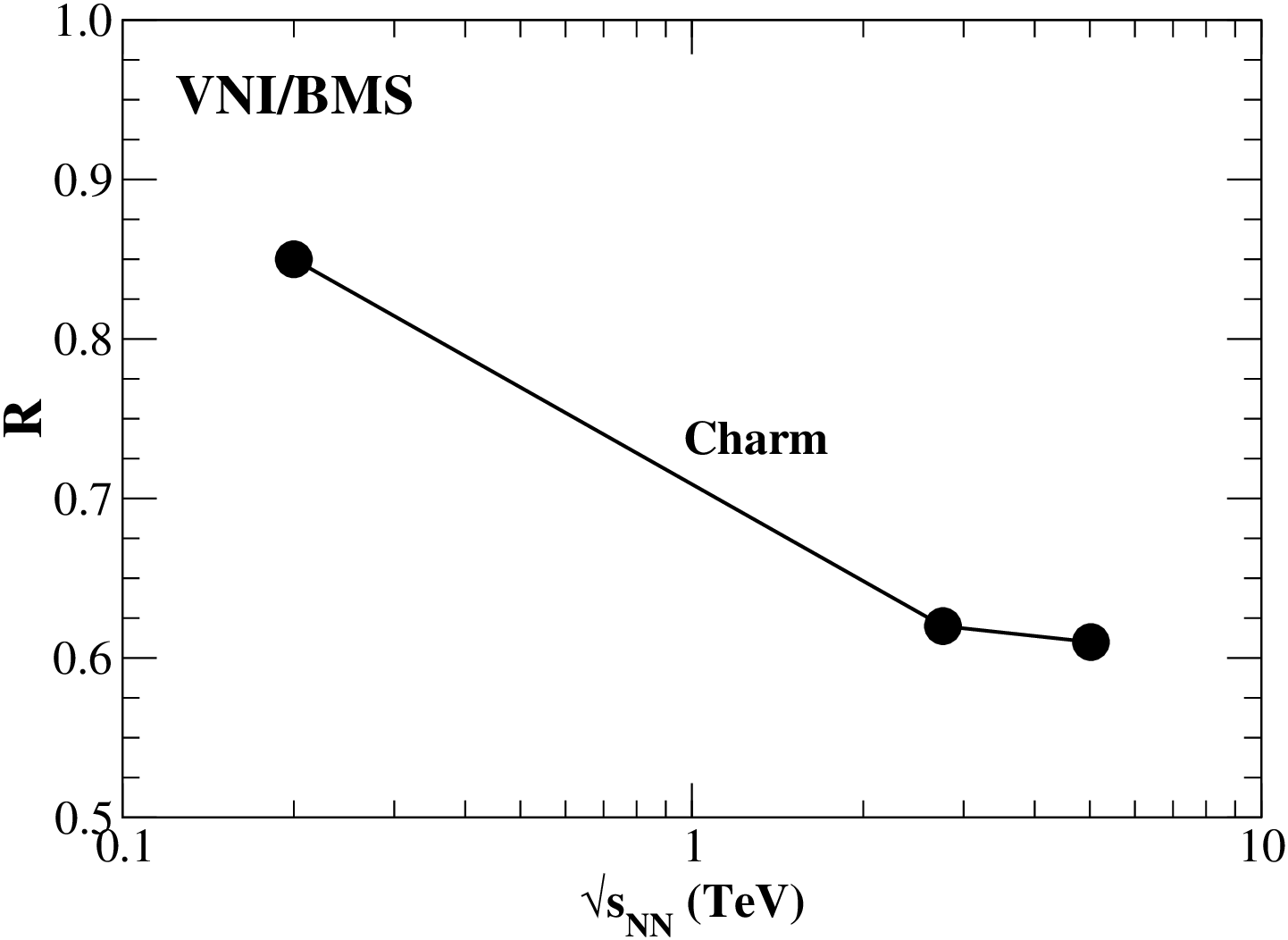}}
\caption{(Colour on-line) The $p_T$ and rapidity integrated production of charm quarks at the end of the pre-equilibrium phase for nucleus-nucleus collisions (for $b=0$ fm) 
and $pp$ collisions scaled by the number of collisions, at the same $\sqrt{s_{\text{NN}}}$. 
 Results are given for Au+Au collisions at 200 AGeV, Pb+Pb collisions at 2.76 ATeV and 5.02 ATeV. The lines are drawn to guide the eye. }
\label{R}
\end{figure}

We shall also use the ratio of $p_T$ and $y$ integrated results to denote the possible deviation of the production
of charm quarks from the $N_{\text{coll}}$ values for $pp$ interactions: 
\begin{equation}
R =\frac{N_\textrm{AA}}{N_\text{coll} \times N_\text{pp}}
\end{equation}

We expect the final results for the medium modification to deviate substantially from what is reported here,
which is only
due to pre-equilibrium dynamics of the system. 

Charm will be conserved during the later evolution of the system. Thus, a rich structure should emerge for the final modification,
once the energy loss suffered by the charm quarks
 and the consequence of the collective flow is accounted for as they traverse the quark gluon plasma.
 The depletion of charm quark at
larger momenta should be accompanied by an enhancement at lower momenta. This enhancement may depend 
strongly on the transverse momentum as the $p_T$ spectrum falls rapidly with increase in $p_T$.
 Further, as charm quarks participate in the collective expansion of the medium, their flow would lead to 
a depletion of charm having very low momenta which would result in an enhancement at intermediate momenta.

In Fig.\ref{pt} we plot the $p_T$ distribution of charm quarks for central $AA$ collisions at $y=0$ along with the same for $pp$ collisions at the corresponding $\sqrt{s_{\text{NN}}}$ scaled by the number of collisions as appropriate for production mechanisms
involving hard interactions. We notice a $p_T$ dependent suppression of charm production in $AA$ collisions, increasing with the
centre of mass energy of the collisions.

The $p_T$ integrated rapidity distributions shown in Fig.~\ref{dy}  brings this fact even more clearly. It additionally suggests that these
modifications are limited to central rapidities at the lowest centre of mass energy (200A GeV) considered here but extend to more forward (backward) rapidities as the energy rises to those available at LHC.

\begin{figure}
\centerline{\includegraphics*[width=8.0 cm]{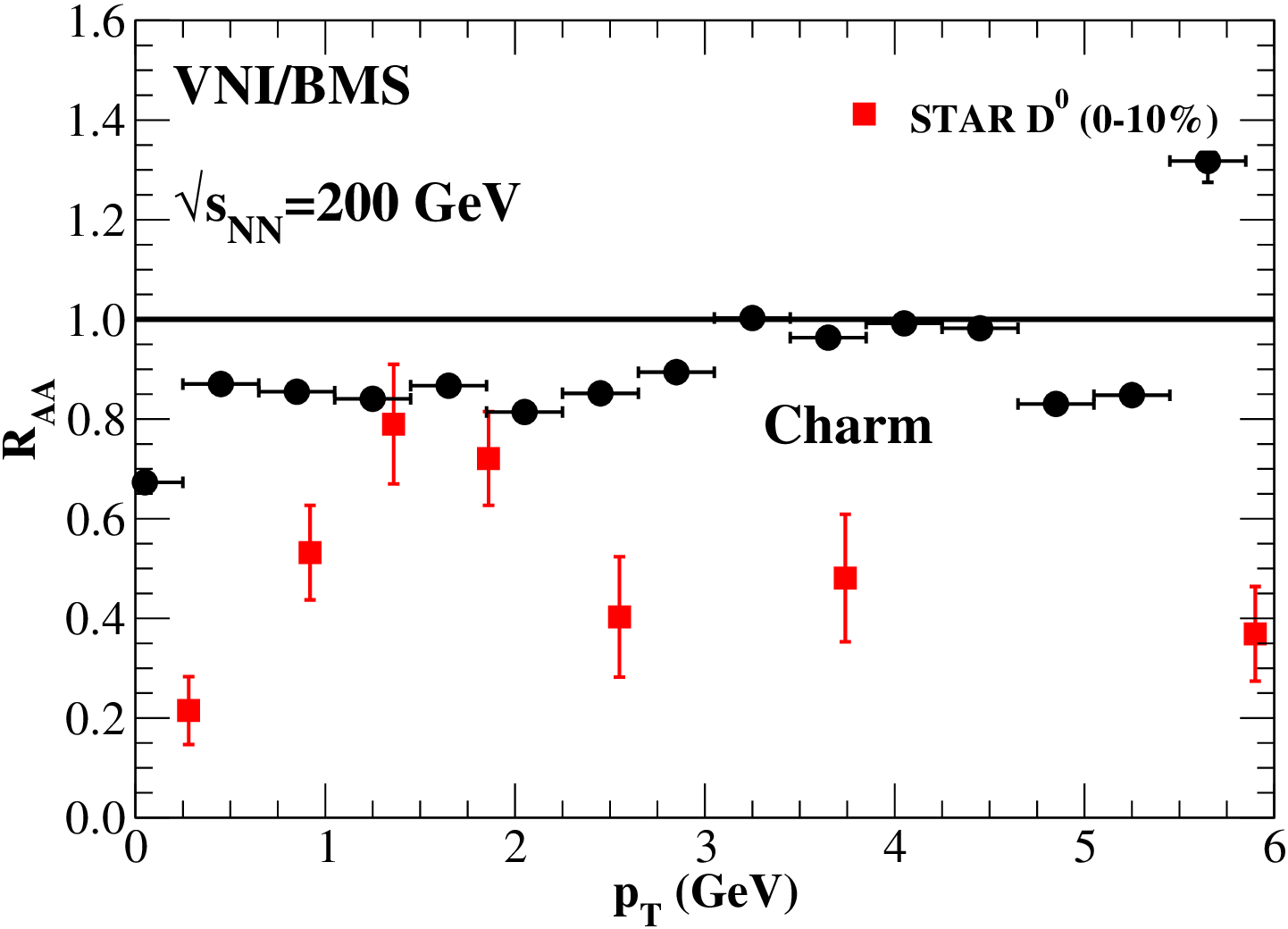}}
\centerline{\includegraphics*[width=8.0 cm]{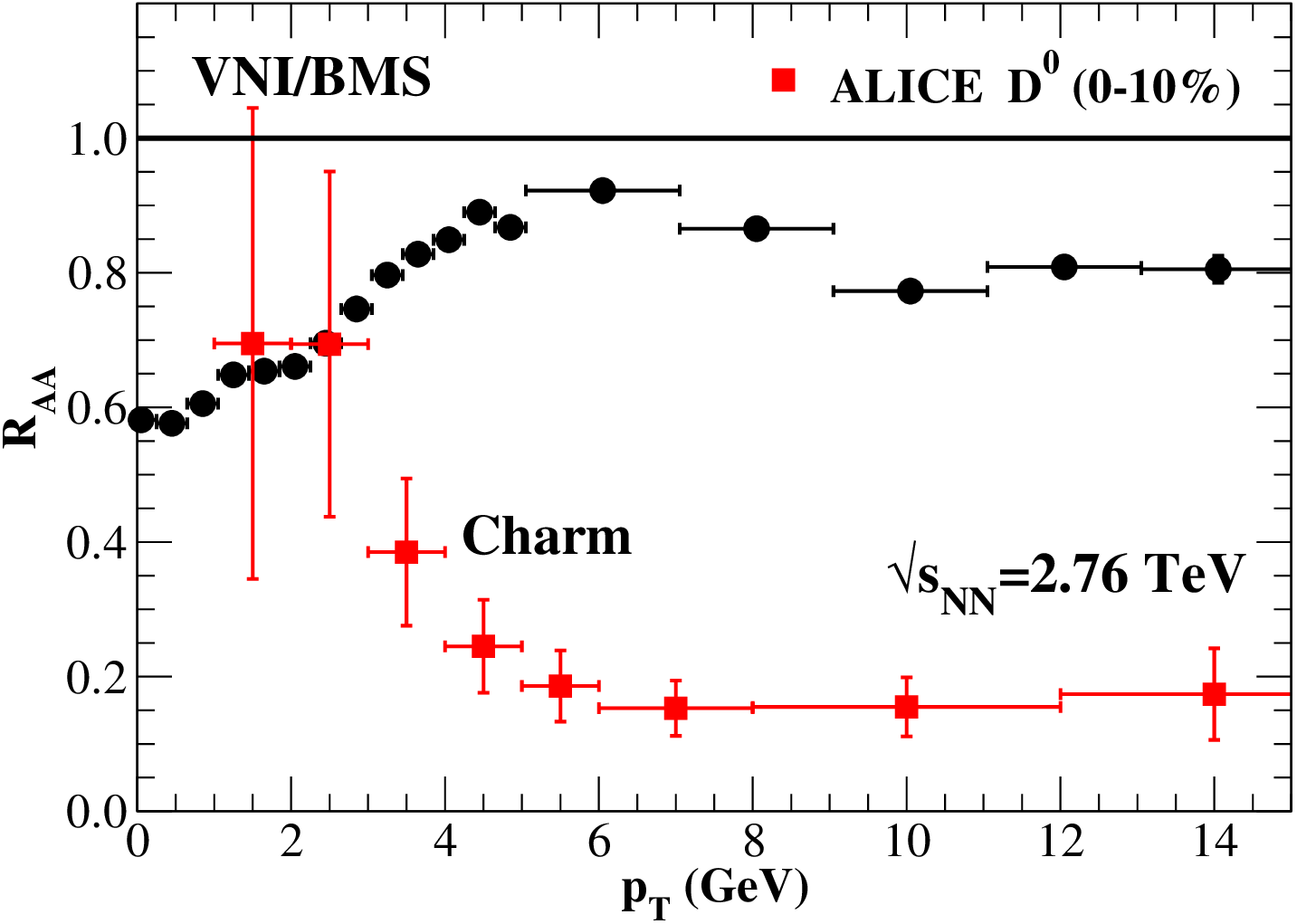}}
\centerline{\includegraphics*[width=8.0 cm]{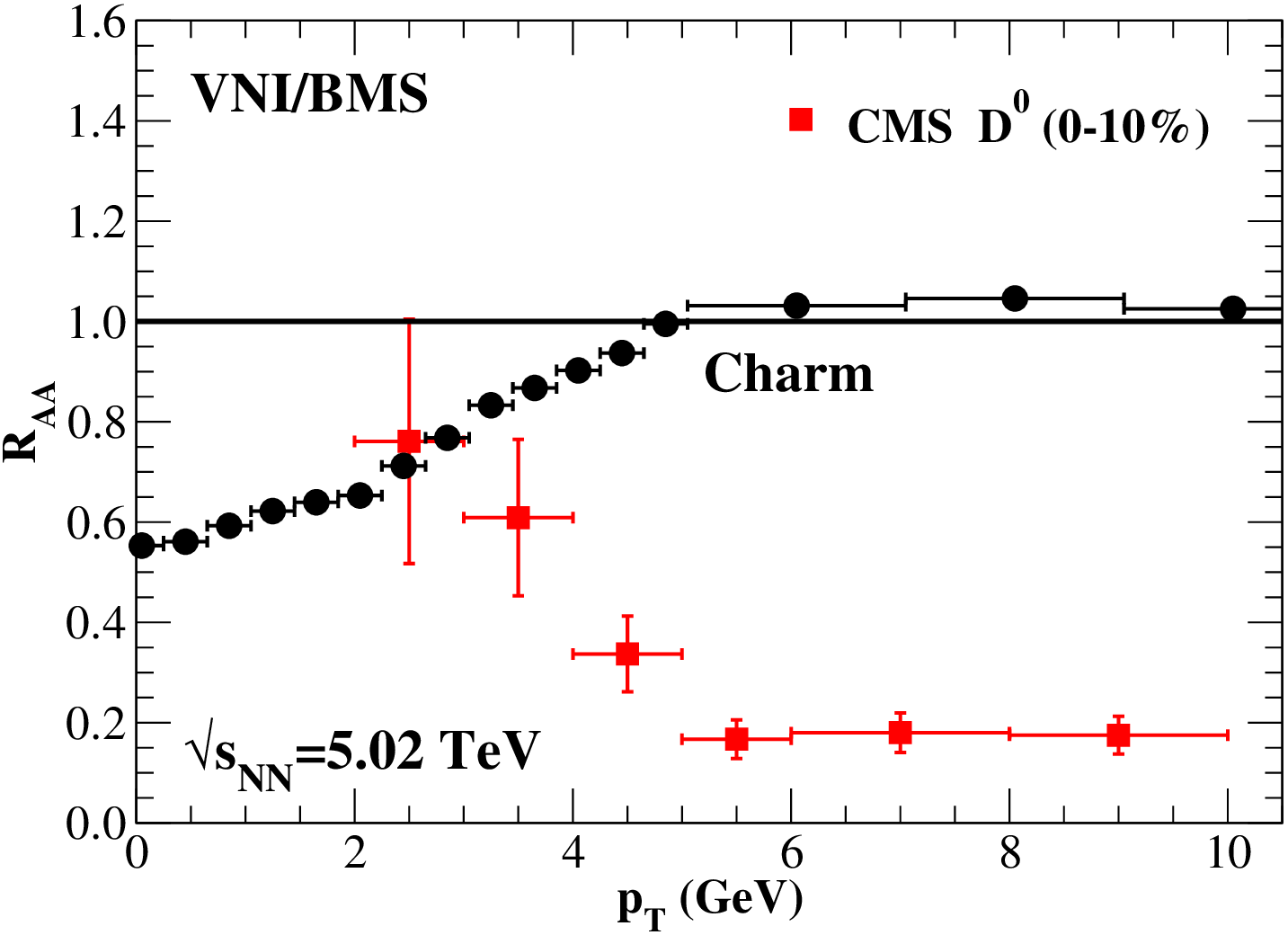}}
\caption{ The medium modification of charm production due to the pre-equilibrium dynamics for Au+Au collisions at 200 AGeV (upper panel), Pb+Pb collisions at 2.76 ATeV (middle panel) and 5.02 ATeV (lower panel) along with experimental data by STAR~\cite{Adamczyk:2014uip,Adam:2018inb}, ALICE~\cite{Adam:2015sza} and CMS~\cite{Sirunyan:2017xss} Collaborations respectively.}
\label{raa}
\end{figure}

The medium modification of total charm production $R$, is shown in Fig.~\ref{R} as a function of centre of mass energy per nucleon. We note that the suppression increases with $\sqrt{s_\text{NN}}$ and tends to saturate at LHC energies. We are not sure that the importance of this
has been sufficiently emphasized in literature. 

Let us discuss this in a little more detail. The experimentally measured $R_\text{AA}$ at 200 AGeV~\cite{Adamczyk:2014uip}, 2.76~ATeV~\cite{Adam:2015sza}, and 5.02 ATeV~\cite{Sirunyan:2017xss} for the central rapidity is always less than one. We know that the charm production during
the thermally equilibrated phase is rather negligible. This trend should persist at larger rapidities.

The authors of Ref.~\cite{Srivastava:2018dye} reported emergence of LPM effect already in $pp$ collisions, signalling the
formation of a dense medium. As stated earlier, it was
found that even with this suppression of scatterings and parton multiplications, there was multiple parton scatterings beyond the primary-primary collisions followed by fragmentations which provided a reasonable explanation to the experimental data.
 In $AA$ collisions the 
LPM effect should be quite strong. This will result in a large scale suppression of partonic collisions as parton multiplication is arrested
due to the
delayed fragmentaions of off-shell partons following scatterings. This should then lead to an overall suppression of charm production beyond
that expected from a straight forward scaling of results of $pp$ collisions by the number of collisions, seen here. It is also expected that
this effect would get stronger as the centre of mass energy rises. We recall that this was not seen in calculations performed with the neglect of
the LPM effect~\cite{Srivastava:2017bcm}, where $R_{\text{AA}} \geq \, 1$  was seen for low $p_T$ (recall also
 that the $p_T$ distributions drop rapidly with 
increasing $p_T$). This implies that in the absence of LPM effect, the parton multiplications and multiple scatterings would lead to a charm production in $AA$ collisions well beyond that obtained from a scaling of the corresponding results for $pp$ collisions. We have verified that it is indeed so, for all the three energies considered here. 

This has one interesting and important consequence. While the final $R_\text{AA}$ will result from a rich interplay of collective flow and
energy loss of the charm quarks, its value at lower $p_T$ would already be quite smaller than unity. An effort to attribute this entire 
suppression due to energy loss during the QGP phase alone would necessarily require a larger value for $dE/dx$.

We give our results for medium modification of charm production at $y=0$ for most central collisions in Fig.~\ref{raa}. We emphasize that we 
neither intend nor expect to explain the experimental $R_\text{AA}$. These are shown 
only to denote how this rich input of medium modification of
charm phase space distribution due the pre-equilibrium dynamics, has to provide the platform for the launch of their journey through the hydrodynamic evolution of the system. During this period they will be subjected to the collective flow and further energy loss.

We do note one interesting and possibly valuable result of these studies. The distribution of charm quarks having large $p_T$ $\geq$ 6 GeV or
so does not seem to be affected strongly by the pre-equilibrium dynamics discussed here.

Thus we feel that a charm distribution whose momenta are sampled from those for $pp$ collisions and the 
points of production distributed according to the nuclear 
thickness function $T_{\text{AA}} (x,y)$  
is not quite adequate as in input for study of charm propagation during the QGP phase of the system produced
in relativistic collision of nuclei.

\section{Summary and Discussion}

We have calculated the dynamics of production of charm quark using parton cascade model, which should 
provide a reasonable description of the parton dynamics, albeit limited to, $p_T \geq p_T^\text{cut-off}$ 
and a reasonably modest virtuality $\geq \mu_0$ defined earlier during the pre-equilibrium phase.
 The LPM effect provides a rich 
structure to the so-called medium modification factor for charm quarks, defined as a ratio of
the results for $AA$ collisions divided by the results for $pp$ collisions (at the same $\sqrt{s_\text{NN}}$)
scaled by number of collisions.We noticed an over all suppression of charm production, which we attribute to the LPM effect.

The medium modification factor as a function of $p_T$ shows a rich structure which evolves with energy, deviating (supression)
 from unity
by about 10\% at low $p_T$ and approaching unity as intermediate $p_T$ at $\sqrt{s_\text{NN}}=$ 200 GeV.
This deviation (suppression) is seen to rise to about 40\% at LHC energies. An interesting result, seems to be the supression 
of large $p_T$ charm at 2.76 TeV, but not at 5.02 TeV, which we are unable to understand.

Realizing that this should form input to the calculations using hydrodynamics and with collisional and radiative energy
loss of charm quarks to determine $dE/dx$, one expect some interesting deviations with those with the neglect of these
suppressions.

A future study, which is under way, will use more modern structure functions, (we have used GRV HO~\cite{Gluck:1994uf} in these preliminary studies) and
account for straight forward corrections like nuclear shadowing, which will further suppress the production of charm quarks
beyond what is reported here. The results for the phase space distribution of charm quarks at the end of the pre-equilibrium 
phase will then be used as inputs
to hydrodynamics based calculations, as indicated above.

In brief, we see that the pre-equilibrium dynamics of parton scattering and fragmentation
along with LPM effect provides a rich structure to the
 production of charm quarks. We suggest that this effect should be taken into account to get a precise value for the energy
loss suffered by charm quarks and the modification of their $p_T$ distributions due to the flow.

\section*{Acknowledgments} 
DKS gratefully acknowledges the grant of Raja Ramanna Fellowship by the Department of Atomic
Energy. We thankfully acknowledge the High Performance Computing Facility of Variable 
Energy Cyclotron Centre Kolkata for all the help, We thank S. A. Bass for a valuable discussion
which triggered this study.

\end{document}